\documentclass[conference]{IEEEtran}

\usepackage{textcomp}
\usepackage{color}
\usepackage{graphicx}
\usepackage{multirow}
\usepackage{tabularx}
\usepackage{listings}
\usepackage{caption}
\usepackage{float}
\usepackage{hyperref}
\def\BibTeX{{\rm B\kern-.05em{\sc i\kern-.025em b}\kern-.08em
    T\kern-.1667em\lower.7ex\hbox{E}\kern-.125emX}}

\newcommand{\src}[1]{\texttt{#1}}

\makeatletter
\newcommand{\linebreakand}{%
  \end{@IEEEauthorhalign}
  \hfill\mbox{}\par
  \mbox{}\hfill\begin{@IEEEauthorhalign}
}
\makeatother

\begin{document}
\title{A Preliminary Study on Using Large Language Models in Software Pentesting}
\author{\IEEEauthorblockN{Kumar Shashwat}
\IEEEauthorblockA{University of South Florida\\
kshashwat@usf.edu}
\and
\IEEEauthorblockN{Francis Hahn}
\IEEEauthorblockA{University of South Florida\\
fhahn@usf.edu}
\and
\IEEEauthorblockN{Xinming Ou}
\IEEEauthorblockA{University of South Florida\\
xou@usf.edu}
\and
\IEEEauthorblockN{Dmitry Goldgof}
\IEEEauthorblockA{University of South Florida\\
  goldgof@usf.edu}
\linebreakand
\IEEEauthorblockN{Lawrence Hall}
\IEEEauthorblockA{University of South Florida\\
lohall@usf.edu}
\and
\IEEEauthorblockN{Jay Ligatti}
\IEEEauthorblockA{University of South Florida\\
ligatti@usf.edu}
\and
\IEEEauthorblockN{S. Raj Rajagopalan}
\IEEEauthorblockA{Resideo\\
siva.rajagopalan@resideo.com}
\and
\IEEEauthorblockN{Armin Ziaie Tabari}
\IEEEauthorblockA{CipherArmor\\
tabari@Cipherarmor.com}
}


\maketitle

\begin{abstract}
  Large language models (LLM) are perceived to offer promising
  potentials for automating security tasks, such as those found in
  security operation centers (SOCs). As a first step towards
  evaluating this perceived potential, we investigate the use of LLMs
  in software pentesting, where the main task is to automatically
  identify software security vulnerabilities in source code. We
  hypothesize that an LLM-based AI agent can be improved over time for
  a specific security task as human operators interact with it. Such
  improvement can be made, as a first step, by engineering prompts fed
  to the LLM based on the responses produced, to include relevant
  contexts and structures so that the model provides more accurate
  results. Such engineering efforts become sustainable if the prompts
  that are engineered to produce better results on current tasks, also
  produce better results on future unknown tasks. To examine this
  hypothesis, we utilize the OWASP Benchmark Project 1.2 which
  contains 2,740 hand-crafted source code test cases containing
  various types of vulnerabilities. We divide the test cases into
  training and testing data, where we engineer the prompts based on
  the training data (only), and evaluate the final system on the
  testing data.  We compare the AI agent's performance on the
  testing data against the performance of the agent without the prompt
  engineering.  We also compare the AI agent's results against those
  from SonarQube, a widely used static code analyzer for security
  testing. We built and tested multiple versions of the AI agent using
  different off-the-shelf LLMs -- Google's Gemini-pro, as well as
  OpenAI's GPT-3.5-Turbo and GPT-4-Turbo (with both chat completion and assistant
  APIs). The results show that using LLMs is a viable approach to
  build an AI agent for software pentesting that can improve through
  repeated use and prompt engineering.
\end{abstract}


\section{Introduction}

Large language models (LLMs) have made massive advancements in
recent years. It has been hoped that LLMs can play a pivotal role in
automating cyber security operations, denting the asymmetric
advantages enjoyed by adversaries. LLMs have demonstrated human-like
reasoning capabilities that are likely useful for analyzing security
events, such as those found in a security operations center (SOC).
Companies are racing to embrace LLMs in security service offerings,
e.g., Microsoft's Security Co-pilot~\footnote{\url{https://www.microsoft.com/en-us/security/business/ai-machine-learning/microsoft-security-copilot}}.
However, there is currently very little information available
regarding how these systems are designed and very little evidence
regarding the effectiveness of using LLMs in the security domain.
Recently, using LLMs in security pentesting has attracted some
interest~\cite{PentestGPT,AIPwnd}.  Using LLMs in pentesting shares
many similarities using LLMs in SOC operations. Both need to
address the large amounts of false alarms, and the ability of
``hunting'' for attacks/vulnerabilities that are not readily reported
by existing tools. The reasoning involved in these security operations
is often nuanced and context-relevant. It is hard to build a
one-size-fit-all tool that can handle all situations, and thus human
involvement is needed for the reasoning to move forward and for
making a final decision. The challenge is that human's brains, while
more capable handling the nuanced situations than a computer
program, are bandwidth-limited and can easily succumb to
burnout~\cite{sundaramurthy2015human} from repeated tasks with similar
structures. Unlike a traditional computer program, an LLM can be
trained on large amounts of data and produce responses to queries
(prompts) that often times demonstrate the type of nuanced reasoning
capability of a human brain. Thus using LLMs in these security tasks
has the potential to automate those tasks that are hard to automate
using traditional computer programs.

In this paper we evaluate the viability of using LLMs in software
pentesting.  In the software development life cycle, pentesting is often
considered one of the last steps~\cite{Culture}. The development team
and the pentesting team often work separately -- remotely or in
different locations, which adds a barrier to communication between
them. Given the workload of a regular pentester it is hard for
them to go through the code files line by line and craft a software
pentesting plan curated just for a specific codebase. They
often end up testing things that are limited to their
information and expertise. Software pentesters use a number of
tools for checking program source code and identifying vulnerabilities,
such as Fortify\footnote{\url{https://www.microfocus.com/en-us/cyberres/application-security/static-code-analyzer}}
and SonarQube\footnote{https://www.sonarsource.com/lp/products/sonarqube/static-code-analysis/}.
These tools often report a large number of findings that turn out to be
false alarms. Large numbers of false alarms lead to pentester fatigue,
and eventually ignoring code analyzer's output altogether.
It would be ideal if these automated tools can ``learn from'' the
pentesters as to why certain findings are false alarms, and use
the learned knowledge to refine future output for the pentesters.
This would only be possible for a traditional computer program if
the developer of the tool is involved in its usage and modify
the tool based on the observed deficiencies. However this is unrealistic
since developers of tools and the tools' users (pentesters) work under
quite different constraints and paces. The feedback loop from users to
developers and back to users (revised tool) is too long to produce
any practical impact. LLMs, on the other hand, can be ``trained'' on the fly
in various ways. One approach is through providing more
prompts that offer the needed knowledge and context, so that the same
LLM model can produce responses that match better with users' expectations.
This may lead to a dynamic AI security agent that can adapt to the specific
usage environment and become more efficient as it interacts with
the human user. 



To evaluate this hypothesis, we built a number of AI agents using
OpenAI's GPT models~\cite{GPT-3.5,GPT-4} and Google's Gemini
model~\cite{Gemini}.  Specifically, we used the LLMs GPT-3.5-Turbo,
GPT-4-Turbo, and Gemini-pro. For the GPT models, we built two agents
for each model, one using the Chat Completions
API\footnote{https://platform.openai.com/docs/guides/text-generation}
and the other using the Assistants
API\footnote{\url{https://platform.openai.com/docs/assistants/overview}}.
We designed prompts for these LLMs and feed the program source code to
them.  We then ask a question to the LLMs about what vulnerabilities are present
in the source code and the location (line number) of the
vulnerability.  We use the test cases published in the OWASP Benchmark
Project\footnote{https://owasp.org/www-project-benchmark/} to evaluate
the accuracy of these AI agents. The benchmark contains 2740 Java
programs with a variety of vulnerabilities such as SQL injection,
cross-site scripting, weak hashing algorithm, and so on.  We compare
the results against SonarQube which is a widely tool used in software
industry for checking software source code for
vulnerabilities. SonarQube also performs better on the OWASP benchmark
than the majority of other static software pentesting tools. To
examine the capability for the AI agent to be improved through prompt
engineering, we divided the benchmark's test cases into training and
testing set. The prompts used in the agents are augmented based on
observing the agents' responses on the training set. The goal of augmenting
the prompts is to add guidance specific to the category of the task 
the LLM is currently trying to accomplish so that higher accuracy can be achieved. 
The new prompts are then tested on the testing set, which has never been seen during
the prompt engineering process. We compare the performance of the AI
agents using the original base prompts, and the agents using the
augmented prompts. We observed the following.

  \begin{enumerate}
  \item Without prompt engineering, the LLMs' accuracy is either below or on par with that of SonarQube.
    
  \item With prompt engineering, GPT-4-Turbo using the Assistants API demonstrated substantial improvements on
    the accuracy, outperforming or being on par with SonarQube in most of the vulnerability categories.
  \end{enumerate}

  These results show that there is a viable path for using LLM to build an AI agent that can be constantly
  improved through prompt engineering driven by usage.  We further compared the
  cases where an LLM model performs differently. The analysis shows that a key reason why LLMs cannot perform
  better is the insufficient understanding of program code flow.

\section{Background}
\subsection{Software Pentesting}
Software pentesting's goal is to identify security vulnerabilities in program code.
It is widely used as part of a company's secure software development life cycle~\cite{microsoft:sdl:book}.
Tools used in software pentesting are divided into two categories: static application security testing (SAST) tools
and dynamic application security testing (DAST) tools. The work described in this paper focuses on SAST only.


\subsection{OWASP Benchmark}

\begin{table}[h]
\centering
\renewcommand{\arraystretch}{1.5}
\begin{tabular}{|c|c|c|c|}
    \hline
    \textbf{Vulnerability Area} & \textbf{True Positive} & \textbf{False Positive} & \textbf{Total} \\
    \hline
    Command Injection   &   126 &   125 &   251 \\
    Weak Cryptography   &   130 &   116 &   246 \\
    Weak Hashing    &   129 &   107 &   236 \\
    LDAP Injection  &   27    &   32  &   59 \\
    Path Traversal  &   133 &   135 &   268 \\
    Secure Cookie Flag	& 36    &   31  & 67 \\
    SQL Injection   &   272 &   232 &   504 \\
    Trust Boundary Violation & 83   &   43  &   126 \\
    Weak Randomness & 218   &   275 &   493 \\
    XPATH Injection & 15    &   20 &    35  \\
    Cross-Site Scripting    & 246   &   209 &   455 \\
    \hline
    \textbf{Total} & \textbf{1415} & \textbf{1325}  &   \textbf{2740} \\
    \hline
\end{tabular}
\vspace{10pt}
\caption{OWASP Benchmark v1.2 Test Cases}
\label{tab:vuln}
\end{table}

The OWASP Benchmark is a Java test suite for evaluating automated software vulnerability detection tools, including both SAST and
DAST. We used the test cases in v1.2, which is a fully executable web application. The benchmark consists of 2740 test cases, each
of which is a separate webpage inside the web app. All the vulnerabilities present in the benchmark are fully exploitable.
The benchmark organizes the test cases based on the type of vulnerability present in the code. Each test case has either
zero or one vulnerability present. Ground truth is given for each test case -- true positive (vulnerability present) or false
positive (vulnerability not present). 
Table~\ref{tab:vuln} shows the distribution of test cases across vulnerability types and ground truth.
 
\subsection{SonarQube}
SonarQube is a widely used SAST tool in industry. In this work we used the
SonarQube Community Edition\footnote{\url{https://docs.sonarsource.com/sonarqube/latest/}}
test results present inside the benchmark and compared them against the LLMs' results.

\subsection{Large Language Models Used}

\begin{table}[h]
\centering
\renewcommand{\arraystretch}{1.5}
\begin{tabularx}{\linewidth}{|X|X|X|X|}
    \hline
    \textbf{Shorthand Name} & \textbf{Model Name} & \textbf{API Used} \\
    \hline
    GPT-3.5-Turbo   &   gpt-3.5-turbo &   ChatCompletion \\
    GPT-4-Turbo   &   gpt-4-1106-preview &   ChatCompletion \\
    Gemini-Pro   &   gemini-pro &   google-generativeai \\
    GPT-3.5-Turbo Assistant   &   gpt-3.5-turbo &   AssistantsAPI \\
    GPT-4-Turbo Assistant   &   gpt-4-1106-preview &   AssistantsAPI \\
    \hline
\end{tabularx}
\vspace{10pt}
\caption{LLMs Used in the Research}
\label{tab:llms}
\end{table}

We used three LLMs: Google's Gemini Pro~\cite{Gemini}, OpenAI's
GPT-3.5-Turbo~\cite{GPT-3.5}, and
GPT-4-Turbo~\cite{GPT-4}. For each GPT model OpenAI provides
two versions of APIs to interact with the models: the Chat Completions API
and the Assistants API. In the paper we use a shorthand name to refer
to a combination of LLM model and API used in the AI agent (Table~\ref{tab:llms}).

\section{Prompt Engineering}

We divide the test cases in the OWASP Benchmark into a training set and a testing set. The division is
done randomly within each vulnerability category, to have 20\% of the test cases in each category in the
training, and the rest in testing. Only the code in the training set was seen in the prompt engineering
process.

\subsection{Base Prompts}

\noindent The goal of the base prompt is to provide just enough guidance and context to the LLM to accurately portray the knowledge and direction an entry level software 
          penetration tester would have when analyzing code. This can be seen in the format of the prompt where it provides a role 

\begin{quotation}
    ``You are a security code analysis tool. Your job is to find security vulnerabilities in the code ...",
\end{quotation}

\noindent it also provides additional mandates that model how one would perform due diligence when working in the field such as, \textbf{``Double check your report."} 
          and \textbf{``Only report something ... if you are 100 percent confident ... "}. We also provide in the base prompt, working directives which explain what to look for and when to report,

\begin{quotation}
    \noindent ``Look at the following code and tell me what vulnerabilities are present in it if any." 
\end{quotation} 

\noindent and at the end of the prompt we go on to provide the types of vulnerabilities present and how to report them. \\

\begin{figure}[htb]
    \centering
    \renewcommand{\arraystretch}{1.5}
    \begin{tabularx}{\linewidth}{|X|}
        \hline
        You are a security code analysis tool. Your job is to find security vulnerabilities in the code with minimum noise. Double check what your report. Only report something as a vulnerability if you are 100 percent confident that it can be exploited now or in the near future. 
        Look the following code and tell me what vulnerabilities are present in it if any. Mention the type of the vulnerability as one of the following: cmdi for Command Injection, crypto for Weak Cryptography, hash for Weak Hashing, ldapi for LDAP Injection, pathtraver for Path Traversal, securecookie for Secure Cookie Flag, sqli for SQL Injection, trustbound for Trust Boundary Violation, weakrand for Weak Randomness, xpathi for XPATH Injection, xss for Cross-site scripting, none for None of these vulnerabilities. \textbf{CODE\_GOES\_HERE}. All output must be in CSV format. You should output the category of the vulnerability from the above mentioned list. The line number of vuln and the reason. Don't output the header for CSV. Eg: weakhash,51,MD5 hash function is used for hashing. MD5 is a weak hashing algorithm.\\
        \hline
    \end{tabularx}
    \vspace{10pt}
    \caption{Base Prompt}
    \label{tab:base_prompt}
\end{figure}

\subsection{LLM Errors on Benchmark Cases under Base Prompts}
\label{subsection:vuln_classification}
After going through the prompt training set, we noticed that the cases where LLMs tend to make
mistakes are false positives and that they can be broadly classified into two types. 

\subsubsection{Code Flow}
\label{subsubsection:vuln_classification_i}

In this type, the program being vulnerable or not depends upon code
flow and the LLM cannot reason about the code flow correctly.
Table~\ref{tab:code_flow_vuln} shows two simplified examples of false
positives from the benchmark. Both were marked incorrectly by
GPT-4-Turbo, and correctly by GPT-4-Turbo Assistant.  Under Benchmark
\#02669, we can see that the value of \src{bar} is always going to be
the string ``\src{safe3}'', thus the user-provided parameter
\src{param} never gets injected in the \src{bar} variable and the code
is not vulnerable. In Benchmark \#007238, we can see the value of
\src{bar} is always going to be the string ``\src{safe}'', and the
user parameter will not be injected.

\begin{table*}[h]
    \centering
    \renewcommand{\arraystretch}{1.5}
\begin{tabularx}{\linewidth}{|X|X|}
    \hline
    \textbf{Pathtraver: Benchmark \#02669} & \textbf{Command Line Injection: Benchmark \#00738} \\ \hline
    \begin{lstlisting}[language=Java, basicstyle=\small, breaklines=true]^^J
        String bar = ``safe1'';^^J
        List<String> valuesList = new ArrayList<>();^^J
        valuesList.add(``safe2'');^^J
        valuesList.add(param);^^J
        valuesList.add(``safe3'');^^J
        valuesList.remove(0);^^J
        bar = valuesList.get(1);^^J
      \end{lstlisting}&
    \begin{lstlisting}[language=Java, basicstyle=\small, breaklines=true]^^J
     String bar;^^J
     int num = 86;^^J
     bar = ((7*42) - num > 200) ? ``safe'' : param;^^J
   \end{lstlisting} \\ \hline
\end{tabularx}
\vspace{10pt}
\caption{Code Flow}
\label{tab:code_flow_vuln}
\end{table*}

\subsubsection{Use of weak algorithms}
In this type the program being vulnerable or not depends upon whether
it uses a weak algorithm, and the LLM fails to determine that the
algorithm is actually not weak.  Table~\ref{tab:weak_algo_vuln} shows
two simplified examples from the benchmark, which again are false
positive. Under Benchmark \#00443, ``\src{AES/GCM/NOPADDING}'' is not
a weak algorithm.  In Benchmark \#00640, the ``\src{getProperty}'' function tries to read the property ``\src{hashAlg2}'' from a file and if the operation fails it falls back to ``\src{SHA-5}''. The value of  ``\src{hashAlg2}'' as stored in the file is SHA-256, not a weak hashing algorithm. Since the LLM is not given the file's content it is unable to determine what hashing algorithm is used. The value of hashAlg2 is supplied in the augmented prompt as shown in Table~\ref{tab:added_prompts}.
\label{subsubsection:vuln_classification_ii}
\begin{table*}[h]
    \centering
    \renewcommand{\arraystretch}{1.5}
\begin{tabularx}{\linewidth}{|X|X|}
    \hline
    \textbf{Weak Cryptography: Benchmark \#00443} & \textbf{Weak Hashing: Benchmark \#00640} \\ \hline
    \begin{lstlisting}[language=Java, basicstyle=\small, breaklines=true]^^J
        javax.crypto.Cipher c = javax.crypto.Cipher.getInstance(``AES/GCM/NOPADDING'')^^J
      \end{lstlisting}&
    \begin{lstlisting}[language=Java, basicstyle=\small, breaklines=true]^^J
        String algorithm = benchmarkprops.getProperty(``hashAlg2'', ``SHA5'');^^J
   \end{lstlisting} \\ \hline
\end{tabularx}
\vspace{10pt}
\caption{Weak Algorithms}
\label{tab:weak_algo_vuln}
\end{table*}

\subsection{Augmenting Prompts}
Each error made by LLM falls into one of the two categories as
discussed above. Prompts are added to correct these errors
based on the category they belong to. Weak Cryptography, Weak Hashing,
and Weak Randomness fall in the ``Use of Weak Algorithms'' category. Command
Injection, LDAP Injection, Path Traversal, Secure Cookie Flag, SQL
Injection, Trust Boundary Violation, XPATH Injection, Cross-site
scripting fall in the ``Code Flow'' category.
The added prompts are listed in Table~\ref{tab:added_prompts}.

\begin{table*}[h]
    \centering
    \renewcommand{\arraystretch}{1.5}
    \begin{tabularx}{\linewidth}{|c|X|}
        \hline
        \textbf{Vulnerability} & \textbf{Prompt} \\
        \hline
        Command Injection   &   Before reporting cmdi, carefully look at the value that is being supplied to arglist variable. If the arglist value contains a constant string not containing the param then there is no cmdi vunerability. \\
        \hline
        Weak Cryptography   &   Only DES/CBC/PKCS5Padding is considered a weak crypto algorithm. cryptoAlg1 is DES/ECB/PKCS5Padding and hashAlg2 is AES/CCM/NoPadding. Consider that benchmark file is always read successfully. \\ 
        \hline
        Weak Hashing    &   Only SHA1 and MD5 are considered weak hashing algorithms. hashAlg1 is MD5 and hashAlg2 is SHA-256. Consider that benchmark file is always read successfully.  \\
        \hline
        LDAP Injection  &   Before reporting ldapi, carefully look at the filter for the ldap client. If the user provided parameter can't be injected into the filter then there is no ldapi security vulnerability. \\
        \hline
        Path Traversal  &  Before reporting pathtraver, carefully look at the bar value that is being injected in the filename variable. If user provided parameter isn't being injected in the filename parameter then there then there is no vulnerability. \\
        \hline
        Secure Cookie Flag	& Before reporting securecookie, carefully look at the bar value that is being supplied to the cookie. If user provided parameter isn't being injected in the cookie then there then there is no securecookie vulnerability.  \\
        \hline
        SQL Injection   &   Before reporting sqli, carefully look at the bar value that is being injected in the sql query. If user provided parameter isn't being injected in the sql query then there then there is no vulnerability. For this codebase, SQL queries without the use of  PreparedStatement can be safe from SQL Injection.  \\
        \hline
        Trust Boundary Violation & Before reporting trustbound, carefully look at the value that is being supplied to request.getSession().putValue(var, "ANY NUMBER"); If the var value contains a constant string not containing the param then there is no vunerability. \\
        \hline
        Weak Randomness & The use of java.util.Random means a weak cryptography vulnerability is present. For this code base the use of java.security.SecureRandom("SHA1PRNG") implies a strong cryptography is used. \\
        \hline
        XPATH Injection & Before reporting xpathi, carefully look at the value that is being supplied to the expression which is fed to nodelist.If the expression value contains a constant string not containing the param then there is no xpathi vunerability.  \\
        \hline
        Cross-Site Scripting    & Before reporting xss, carefully look at the bar variable that is specified to response.getWriter function. If the bar variable contains a constant string not containing the param then there is no xss vunerability. \\
        \hline
    \end{tabularx}
    \vspace{10pt}
    \caption{Added Prompts}
    \label{tab:added_prompts}
\end{table*}


\section{Experimentation and Evaluation}

For evaluation and experimentation purposes, we used the OWASP
software testing suite version 1.2. The suite contains 2740 source
files designed with a single vulnerability from the 11 categories
as listed in Table~\ref{tab:vuln}. In order to generate the
augmented prompts for each vulnerability, we divided the dataset into
a 20:80 split of the entire set of data. We only looked at the 20\% of
the source files to generate the augmented prompts and tested the
performance of those prompts on the 80\% of the data. 
This experimentation strategy models a real-world scenario where a pentester would look
at the pentesting tool's result and understand some reported findings are
false positives. The pentester then extrapolates the
causes of the mistake and provide additional guidance to the LLM in the form
of added prompts.
Next time when a new program is analyzed, the augmented prompts avoid
making the same errors.
In our study two types of experiments were performed. 

\begin{enumerate}
    \item Our first experiment was performed using Figure~\ref{tab:base_prompt} as base prompt with limited information about the context of the types of vulnerabilities present. We only provided the 
          categories of vulnerabilities to ensure that the formatting of the LLM's output fits the scoring engine.
    \item For the second experiment we appended the added prompt from Table~\ref{tab:added_prompts} for each vulnerability category. The augmented prompts contained specific detailed guidance pertaining to each category, based on the observation from the training data.
\end{enumerate}

The augmented prompts provide more context to the base prompt by
telling the LLM what is considered a vulnerability with respect to the
codebase. For example: under Weak Hashing where we direct the LLM to
consider only SHA1 and MD5 to be weak hashing algorithms, variables
such as \src{hashAlg1} and \src{hashAlg2} are to be MD5 and SHA-256
respectively.


\begin{table*}[htb]
  \centering
  \renewcommand{\arraystretch}{1.5}
  \begin{tabularx}{\linewidth}{|p{80pt}|X|X|X|X|X|X|X|}
      \hline
      \textbf{Vulnerability} & \textbf{SonarQube} & \textbf{Prompt} & \textbf{GPT-3.5-Turbo} & \textbf{GPT-4-Turbo} & \textbf{Gemini Pro} & \textbf{GPT-3.5-Turbo Assistant} & \textbf{GPT-4-Turbo Assistant}\\ \hline

      \multirow{2}{*}{Command Line Injection} & \multirow{2}{*}{49.8\%} & Base& 38.2\% &49.2\% &50.2\% &53.8\% & 70.3\% \\ & & Augmented& 49.2\% &47.7\% &50.2\% &50.2\% & 74.3\% \\ \hline

      \multirow{2}{*}{Weak Cryptography} & \multirow{2}{*}{89.0\%} & Base& 28.0\% &50.0\% &53.0\% &46.5\% & 74.5\% \\ & & Augmented& 53.0\% &52.5\% &53.5\% &54.5\% & 89.7\% \\ \hline

      \multirow{2}{*}{Weak Hashing} & \multirow{2}{*}{83.0\%} & Base& 32.6\% &51.5\% &32.9\% &44.5\% & 71.8\% \\ & & Augmented& 54.2\% &55.3\% &53.7\% &50.0\% & 85.1\% \\ \hline

      \multirow{2}{*}{LDAP Injection} & \multirow{2}{*}{54.2\%} & Base& 11.8\% &42.5\% &44.6\% &53.1\% & 51.0\% \\ & & Augmented& 42.5\% &40.4\% &44.6\% &51.0\% & 57.4\% \\ \hline

      \multirow{2}{*}{Path Traversal} & \multirow{2}{*}{100\%} & Base& 50.3\% &48.5\% &50.0\% &56.7\% & 62.6\% \\ & & Augmented& 49.0\% &47.6\% &49.5\% &53.0\% & 70.5\% \\ \hline

      \multirow{2}{*}{Secure Cookie Flag} & \multirow{2}{*}{46.2\%} & Base& 46.2\% &52.8\% &56.6\% &64.5\% & 94.3\% \\ & & Augmented& 54.7\% &52.8\% &54.7\% &41.1\% & 84.9\% \\ \hline

      \multirow{2}{*}{SQL Injection} & \multirow{2}{*}{50.4\%} & Base& 52.7\% &53.9\% &54.4\% &51.0\% & 62.4\% \\ & & Augmented& 50.7\% &51.4\% &54.9\% &45.0\% & 67.8\% \\ \hline

      \multirow{2}{*}{Trust Boundary Violation} & \multirow{2}{*}{34.1\%} & Base& 34.1\% &54.0\% &71.0\% &45.0\% & 56.0\% \\ & & Augmented& 61.0\% &66.0\% &70.0\% &42.1\% & 53.0\% \\ \hline

      \multirow{2}{*}{Weak Randomness} & \multirow{2}{*}{100\%} & Base& 44.8\% &39.6\% &43.0\% &55.4\% & 93.1\% \\ & & Augmented& 40.9\% &40.9\% &42.7\% &47.2\% & 98.7\% \\ \hline

      \multirow{2}{*}{XPATH Injection} & \multirow{2}{*}{57.1\%} & Base& 45.7\% &40.7\% &40.7\% &33.3\% & 59.2\% \\ & & Augmented& 45.7\% &40.7\% &40.7\% &14.8\% & 74.0\% \\ \hline

      \multirow{2}{*}{Cross-Site Scripting} & \multirow{2}{*}{45.9\%} & Base& 45.4\% &50.6\% &58.4\% &52.1\% & 78.7\% \\ & & Augmented& 50.1\% &49.5\% &55.0\% &53.6\% & 76.0\% \\ \hline
  \end{tabularx}
  \vspace{10pt}
  \caption{Experimentation Results}
  \label{tab:results}
\end{table*}



We compare the various LLM models' performance along side with the performance
of SonarQube, an open-source platform used for continuous code inspection and analysis.
All LLM models are provided the same base prompt and augmented prompts.
In Table~\ref{tab:results}, the accuracy percentage is calculated by total number of
correctly predicted cases (either true positive or false positive) divided by the
total number of cases {\it on the testing data}.
The results show that for the GPT-4-Turbo model using the Assistants API, the accuracy
of the AI agent outperforms that of SonarQube under the augmented prompts, for most of
the vulnerability categories. We also see a consistent improvement of accuracy under
the augmented model over the base model, for this combination of LLM model and API.
This result indicates that GPT-4-Turbo using Assistants API provides a viable path
towards using LLMs in software pentesting. In the next section we provide more detailed
discussions on the results.



\section{Discussion}
As shown in Table \ref{tab:results}, we can see that the augmented
prompts do not always increase performace. However, the augmented
prompts perform better for at least one LLM in each category. 
We rate each LLM based on two criterias:
\begin{enumerate}
    \item Ability to learn from augmented prompts
    \item Overall performance in each category
\end{enumerate}

\subsection{GPT-3.5-Turbo}

GPT-3.5-Turbo with ChatCompletion generally had the poorest accuracy
compared with the other LLMs for the base prompt. It showed a
significant jump in performance with augmented prompts in most
categories. However, the augmented prompts did not yield better results
for Path Traversal, SQL Injection, Weak Randomness, and XPATH
Injection.

\subsection{GPT-4-Turbo}
GPT-4-Turbo with ChatCompletion showed a noticable increase in
performance from GPT-3.5-Turbo in all categories except for Weak
Randomness among the base prompts. Performance for the augmented
prompts out performed SonarQube but stayed relatively within the same
performance range as the augmented prompts of GPT-3.5-Turbo.

\subsection{Gemini-Pro}
Gemini-Pro showed consistent performance between the base and
augmented prompts for most categories and matches the capabilities of
the GPT-3.5-Turbo and GPT-4-Turbo models with ChatCompletion. It is
also noted that Gemini-Pro had the highest performance among all of
the experiments in the Trustboundary category with 71\% accuracy for
the base prompt and 70\% accuracy for the augmented prompts.

\subsection{GPT-3.5-Turbo-Assistant}
GPT-3.5-Turbo with the Assistant API showed similar results to GPT-3.5-Turbo 
with ChatCompletion and Gemini-Pro. However, there were
a few instances where the base prompts outperformed all previous
tests. The augmented prompts showed a similar behavior as with the
previous models, but overall increased performance was seen with
this model and API pairing. However, with this experiment we saw a
unique occurence where the augmented prompt had three cases of lower
performance in the augmented prompts, in particular for the SQL
Injection, Weak Randomness, and XPATH Injection categories.

\subsection{GPT-4-Turbo-Assistant}
GPT-4-Turbo with the Assistant API showed the best performance among all of the LLMs and API pairings, aside from Trustboundary where Gemini-Pro performed the best in testing. The base prompts showed a significant increase in performance across all categories aside from Trustboundary and LDAP injections which had comparable performance to the GPT-3.5-Turbo and Assistant API pairing. The augmented prompts showed similar behavior to all other experiments with regards to showing improvements to performance from base to augmented prompts. This came with an exception Secure Cookie Flag category where the GPT-4-Turbo with Assistant API showed similar results of lower performance in the augmented prompts as with the GPT-3.5-Turbo and Assistant API pairing.

\subsection{On Evaluation Strategy}
In our evaluation we used the same prompts for all the LLMs. In reality it makes more sense to 
adopt a more tailored approach, where prompts are engineered based on the specific LLM's responses
and the improvements seen. A single one-size-fit-all process for prompt engineering, while removing
human bias in the evaluation process, does not reflect how LLMs are used and tailored.
A more human-centered approach for evaluation could potentially address this limitation.


\section{Related Work}
Deng et al.~\cite{PentestGPT} presented PentestGPT, an LLM-based
AI agent to faciliate penetration testing. The authors created separate
GPT sessions focusing on macroscopic and microscopic sub tasks to address
the memory loss problem. It also adopts attack trees to guide the multiple
GPT sessions towards the goals of the pentesting. PentestGPT does not
address the question of whether the engineered prompts can be effective
on new pentesting tasks that have not been seen before. 
Happe and Cito~\cite{AIPwnd} discussed the vision of using LLMs in pentesting.
A prototype AgentGPT was constructed that can help a pentester elevate
privilege on a local host. There is no systematic study on the effectiveness
of AgentGPT and no details were given about the prompts used or the prompt
engineering process. In addition to presenting a vision of using LLM in
software pentesting, our work conducted a preliminary study on the efficacy
of LLM in this domain, through experimentations on a well established benchmark.
Our use of prompt engineering is
similar to the work by Espeje et al.~\cite{Prompts}
which discusses various methods of prompt engineering and how they can
be used to improve the performance of LLMs by categorizing the prompts
in various formats and then augmenting original proposals with higher
performing prompts to test the extent of the generation cabilities of
the LLMs. They use these methods of prompt engineering to test the
performance of the LLMs abilities for inductive reasoning, deductive
reasoning, mathematical reasoning, and multi-hop reasoning.


\section{Conclusion}
We present preliminary experimentation study on using LLMs
in software pentesting. Our results show that through prompt
engineering, an LLM can improve its accuracy over usage, and
its accuracy is on par or surpassed SonarQube, a widely used
static software pentesting. 



\section*{Acknowledgment}

This work was
partially supported by the National Science
Foundation under award no. 2235102, and Office of Naval Research under
award no. N00014-23-1-2538. Any opinions, findings and
conclusions or recommendations expressed in this material are those of
the authors and do not necessarily reflect the views of the above funding agencies.



\bibliographystyle{plain}
\bibliography{software_pentest_llms,anth,secdev}

\begin{thebibliography}{1}

\bibitem{GPT-3.5}
Tom~B. Brown, Benjamin Mann, and et~al.
\newblock Language models are few-shot learners.
\newblock {\em arXiv}, 2023.

\bibitem{PentestGPT}
Gelei Deng, Yi~Liu, Víctor Mayoral-Vilches, Peng Liu, Yuekang Li, Yuan Xu,
  Tianwei Zhang, Yang Liu, Martin Pinzger, and Stefan Rass.
\newblock Pentest{GPT}: An {LLM}-empowered automatic penetration testing tool.
\newblock {\em arXiv}, 2023.

\bibitem{AIPwnd}
Andreas Happe and J\"{u}rgen Cito.
\newblock Getting pwn’d by {AI}: Penetration testing with large language
  models.
\newblock In {\em Proceedings of the 31st ACM Joint European Software
  Engineering Conference and Symposium on the Foundations of Software
  Engineering}, ESEC/FSE 2023, page 2082–2086, San Francisco, CA, USA, 2023.

\bibitem{microsoft:sdl:book}
Michael Howard and Steve Lipner.
\newblock {\em The security development Lifecycle}, volume~8.
\newblock Microsoft Press Redmond, 2006.

\bibitem{Prompts}
Jessica {López Espejel}, El~Hassane Ettifouri, Mahaman~Sanoussi {Yahaya
  Alassan}, El~Mehdi Chouham, and Walid Dahhane.
\newblock {GPT}-3.5, {GPT}-4, or {BARD}? evaluating {LLMs} reasoning ability in
  zero-shot setting and performance boosting through prompts, 2023.

\bibitem{GPT-4}
OpenAI*.
\newblock {GPT-4} technical report.
\newblock {\em arXiv}, 2023.

\bibitem{sundaramurthy2015human}
Sathya~Chandran Sundaramurthy, Alexandru~G Bardas, Jacob Case, Xinming Ou,
  Michael Wesch, John McHugh, and S~Raj Rajagopalan.
\newblock A human capital model for mitigating security analyst burnout.
\newblock In {\em Eleventh Symposium On Usable Privacy and Security (SOUPS
  2015)}, pages 347--359, 2015.

\bibitem{Gemini}
Google~Gemini Team.
\newblock {Gemini}: A family of highly capable multimodal models.
\newblock {\em arXiv}, 2023.

\bibitem{Culture}
Anwesh Tuladhar, Daniel Lende, Jay Ligatti, and Xinming Ou.
\newblock An analysis of the role of situated learning in starting a security
  culture in a software company.
\newblock {\em USENIX}, 2021.

\end{thebibliography}

\end{document}